\documentclass[aps,pra,twocolumn,showpacs,groupedaddress,letterpaper]{revtex4-1}

\usepackage{graphicx}
\usepackage{subfigure}
\usepackage{color}
\usepackage{physymb}
\usepackage{enumitem}
\usepackage{mdframed}
\usepackage{framed}
\usepackage{epstopdf}

\begin{document}

\title{Coupled Multiple-response vs. Free-response Conceptual Assessment: \\An Example from Upper-division Physics}

\pacs{01.40.Fk, 01.40.G-, 01.40.gf, 01.50.Kw}

\author{Bethany R. Wilcox}
\affiliation{Department of Physics, University of Colorado, 390 UCB, Boulder, CO 80309}

\author{Steven J. Pollock}
\affiliation{Department of Physics, University of Colorado, 390 UCB, Boulder, CO 80309}

\begin{abstract}
Free-response research-based assessments, like the Colorado Upper-division Electrostatics Diagnostic (CUE), provide rich, fine-grained information about students' reasoning.  However, because of the difficulties inherent in scoring these assessments, the majority of the large-scale conceptual assessments in physics are multiple-choice.   To increase the scalability and usability of the CUE, we set out to create a new version of the assessment that preserves the insights afforded by a free-response format while exploiting the logistical advantages of a multiple-choice assessment.  We used our extensive database of responses to the free-response CUE to construct distractors for a new version where students can select multiple responses and receive partial credit based on the accuracy and consistency of their selections.  Here, we describe the development of this modified CUE format, which we call coupled multiple-response (CMR), and present data from direct comparisons of both versions.  We find that the two formats have the same average score and perform similarly on multiple measures of validity and reliability, suggesting that the new version is a potentially viable alternative to the original CUE for the purpose of large-scale research-based assessment.  We also compare the details of student responses on each of the two versions.  While the CMR version does not capture the full scope of potential student responses, nearly three-quarters of our students' responses to the free-response version contained one or more elements that matched options provided on the CMR version.  
\end{abstract}

\maketitle

\section{Introduction \& Motivation}

Over the past several decades, large-scale conceptual assessments, like the Force Concept Inventory (FCI) \cite{hestenes1992fci} and Brief Electricity and Magnetism Assessment (BEMA) \cite{ding2006bema}, have become a standard part of assessing student learning gains in introductory physics courses in many undergraduate physics departments.  These assessments offer a consistent measure of student learning that can be compared across courses or institutions and can provide data on student difficulties that persist even after instruction \cite{meltzer2012resource}.  Historically, scores on these assessments have served as useful tools for inspiring course transformation efforts aimed, in part, at increasing students' conceptual learning gains in introductory courses \cite{hake1998ie}.  

Despite the value of conceptual surveys like the FCI and BEMA at the introductory level, the use of standardized conceptual assessment is not as common in upper-division courses, where the more advanced physics content poses unique challenges.  These challenges include the necessary use of specialized language and the need for sophisticated mathematical techniques.  However, despite these challenges, several conceptual assessments have been developed for upper-division physics \cite{baily2012materials, caballero2013ccmi, goldhaber2009qmat,singh2006scge,rhoads1999wci,zhu2012qms}, including the Colorado Upper-division Electrostatics (CUE) diagnostic \cite{chasteen2012cue}. The CUE was developed at the University of Colorado Boulder (CU) in conjunction with the development of transformed course materials for junior-level electrostatics \cite{chasteen2012transforming}.  Unlike its introductory counterparts that are composed of multiple-choice questions, the CUE was intentionally designed as a free-response (FR) instrument.  The developers chose a FR format in part because learning goals developed through collaboration with CU faculty emphasized the importance of students synthesizing and generating responses \cite{chasteen2012transforming}.  It was felt that a FR format would more adequately test these consensus learning goals, and thus would be valued by the faculty more than a multiple-choice instrument.  Additionally, relatively little literature on student difficulties around problems related to upper-division electrostatics was available to inform the development of tempting multiple-choice distractors.  The developers anticipated that once established, the FR format might provide the insight into students' reasoning necessary to craft a multiple-choice version of the assessment at a later date.

Since its development, the CUE has been given in multiple courses and institutions \cite{chasteen2012cue}.  CUE scores correlate strongly with other measures of student learning, such as overall course and BEMA score, and are sensitive to different types of instruction (e.g., interactive vs. traditional lecture).  However, grading the CUE requires a complex rubric, and significant training is needed for graders to produce scores that can be compared to other institutions.  The need for training severely limits the CUE's potential as a large-scale assessment tool like the multiple-choice (MC) instruments used at the introductory level.  If the CUE is to be used as a tool by a wide range of faculty, it must be adapted to a more easily graded format without sacrificing its ability to provide a meaningful measure of students' conceptual understanding in upper-division electrostatics.  

To address these scalability and usability issues, we crafted a multiple-choice version of the CUE using student solutions from previous semesters to construct distractors.  Early attempts quickly showed that a standard MC assessment, with a single unambiguously correct answer and 3-5 clearly incorrect (though tempting) distractors, was insufficient to capture the variation in responses found on the majority of the FR questions.  Instead, we developed a version where students select multiple responses and receive partial credit depending on the accuracy and consistency of their selections in order to match the more nuanced grading of a FR assessment \cite{lin2013mcVfr}.  

The purpose of this paper is to present a direct comparison of the new multiple-response format of the CUE with the original FR format.  To do this, we review some of the existing literature on multiple-choice testing (Sec.\ \ref{sec:Lit}) and describe the development, validation, and scoring of this new coupled multiple-response (CMR) CUE (Sec.\ \ref{sec:Development}).  We then present findings from a direct comparison of test statistics from both the FR and CMR formats that show the two formats perform similarly on multiple measures of validity and reliability (Sec.\ \ref{sec:Results}).  We also compare the details of student reasoning accessed by each version of the CUE (Sec.\ \ref{sec:Reasoning}) and conclude with a discussion of limitations and ongoing work (Sec.\ \ref{sec:Discussion}).  The large-scale validation of the CMR CUE as an independent instrument will be the focus of a future publication.

\section{\label{sec:Lit}Background on Multiple-choice testing}

Instructors and researchers have been considering the relative advantages of multiple-choice (MC) and free-response (FR) testing formats for many years \cite{bork1984noMC, sandin1985noMC, scott1985defenseMC, iona1986defenseMC}.  While the difficulties inherent in writing and interpreting MC questions have been well documented \cite{iona1984wordingMC, varney1984writingMC, dufresne2002interpretingMC}, the ease and objectivity of grading afforded by MC formats makes them ideal for large-scale and cross-institutional testing.  These logistical advantages have motivated considerable work to develop strategies for constructing valid and objective multiple-choice items \cite{aubrecht1983objectiveMC, hudson1981constructingMC, sobolewski1996developingMC}.  This includes discussion of how many MC distractors should be used \cite{rodriguez2005MCdistractors} and methods for accounting for guessing \cite{higbie1993guessingMC}.

MC conceptual inventories have now been developed for a wide range of topics in introductory physics (see Ref.\ \cite{hestenes1992fci, thornton1998fmce, ding2006bema} for examples and Ref.\ \cite{beichnerAssessment} for a more comprehensive list).  The items and distractors on these assessments are based on research on student difficulties.  The validity and reliability of these assessments has typically been established using Classical Test Theory (CTT) \cite{engelhardt2009ctt}.  CTT posits specific characteristics of high quality MC tests and provides a number of strategies and statistical measures to determine if an instrument meets these criteria.  The specific requirements and tests of CTT will be discussed in greater detail in Sec.\ \ref{sec:Results}.  

Another potential lens for establishing the validity and reliability of standardized assessments is Item Response Theory (IRT).  IRT involves the estimation of item characteristic parameters and students' latent abilities \cite{ding2009mcanalysis}.  Unlike the test statistics produced using CTT, IRT's estimates of the student and item parameters are test and population independent when the assumptions of the theory are satisfied.  The use of IRT to look at conceptual assessments targeted at undergraduate physics is less common than the use of CTT, but has been done \cite{ding2014bemaVrasch, aslanides2013rciVirt, marshall2009taksVirt, planinic2010fciVrasch}.  Despite the appeal of generating population-independent parameters, the analysis in this paper will exclusively utilize CTT.  This choice was dictated primarily by our goal to replicate an existing conceptual assessment that was neither developed nor analyzed using IRT.  Additionally, even simple dichotomous IRT models require a minimum N of roughly 100 to produce reliable estimates of the item and student parameters \cite{ding2009mcanalysis}, and this minimum number increases when using more complex models or polytomous scoring \cite{deayala2009polyIRT}.  

The increase in the number and usage of MC conceptual inventories has not, however, marked the end of the tension between MC and FR testing formats, and direct comparisons of the two have yielded mixed results.  Kruglak \cite{kruglak1965MCvEssay} looked at tests of physics terminology with both MC and FR formats.  He found that while the MC questions showed a slightly higher correlation with overall achievement in the course than the FR questions, scores on the two different formats showed only low to moderate correlation with one another.  This finding disagrees with results from Hudson and Hudson \cite{hudson1981constructingMC}, who found a high degree of correlation between students' scores on MC tests and a set of FR pop-tests particularly when using an aggregate score from all the MC tests given in the semester.  A related study by Scott et. al. \cite{scott2006MCvFR} compared students' scores on a MC final exam to their performance on the same problems in a think-aloud interview setting.  Ranking the students based on each of these scores, they found a high correlation between the two ranks.  

Additional work comparing MC and FR formats has focused specifically on the FCI.  For example, Lin and Singh \cite{lin2013mcVfr} looked at two FCI questions in MC and FR formats.  As with all FCI items, the distractors for these questions were based on common student difficulties.  However, rather than being scored dichotomously,  scores were weighted to reflect different levels of understanding.  Comparing scores on MC and FR versions of these questions, they found that average scores on the two formats did not differ significantly, and that both formats had equivalent discrimination.  A similar study reported a moderate correlation between student performance on FCI questions and FR exam questions targeting the same physics content \cite{steinberg1997fciMCvFR}.  However, this study also found that the nature of individual student's responses varied significantly between the FCI and exam questions.  

There have also been attempts to narrow the gap between MC and FR formats by modifying the `standard' MC format characterized by a single unambiguously correct answer and 3-5 distractors.  For example, a conceptual inventory developed by Rhoads and Roedel \cite{rhoads1999wci} contained a number of items with a `multiple-correct' answer format.  These items offered four response options including 2-3 correct options targeting different levels of Bloom's Taxonomy \cite{bloom1956taxonomy} in order to differentiate between higher and lower level learning.  Kubinger and Gotschall \cite{kubinger2007multipleTF} also described a multiple-response format where each item has five response options, of which any number may be correct.  Students were scored as having mastered the item only if they selected all correct responses and none of the incorrect ones.  This study found that measures of item difficulty did not differ significantly between the multiple-response format and FR formats.  

Overall, previous research on MC testing has highlighted a number of similarities and differences between MC and FR formats.  Agreement between the two formats appears augmented when: the MC distractors are based on known student difficulties, the scoring of each MC distractor is adjusted to reflect varying levels of student understanding, the format of the MC items is adjusted to reduce the impact of guessing, and/or a large number of MC items are used in the comparison.  We specifically leveraged the first three of these heuristics in our development of the new CMR CUE.

\section{\label{sec:Development}Developing The Coupled Multiple-response CUE}

\subsection{\label{sec:Adapting}Adapting the Questions} 

\begin{figure}
  \begin{minipage}{.93\linewidth}
   \flushleft {\emph{Give a brief outline of the EASIEST method the you would use to solve the problem.  \\ {\bf DO NOT SOLVE the problem, we just want to know: \\ \hspace{2mm}(1) The general strategy (half credit) and \\ \hspace{2mm}(2) Why you chose that method (half credit)}}}  \vspace{2mm}
  \end{minipage}
  \begin{minipage}{.65\linewidth}
      \flushleft {\bf Q3} - \emph{A solid, neutral, non-conducting cube as in the figure, with side length `a' and $\rho(z)=kz$. \\ \vspace{1mm}Find $\vec{E}$ (or V) outside, at point P, where P is {\bf off-axis}, at a distance {\bf 50a} from the cube.}
  \end{minipage}
  \begin{minipage}{.27\linewidth}
     \begin{center}
        \includegraphics[width=19mm, height=25mm]{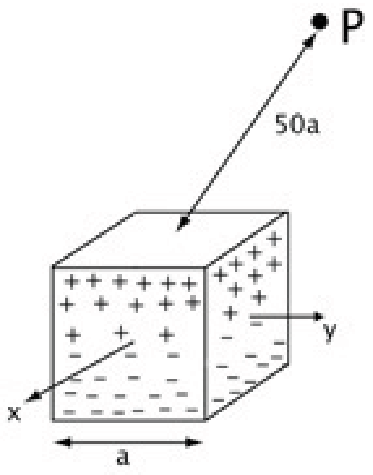}
      \end{center}
  \end{minipage}
\caption{A sample item from the original FR CUE. Question prompt has been paraphrased; see Ref.\ \cite{website} for full question prompt, rubric, and scoring materials.} \label{fig:FRitem}
\end{figure}

As our goal was to create a new version of an already established and validated conceptual assessment, we explicitly avoided making substantive changes to the content coverage or questions on the CUE.  There are several distinct question styles on the original FR CUE, but roughly two-thirds of the items present students with a physics problem and ask them to state the appropriate solution method or select an answer and then to justify that answer/method (Fig.\ \ref{fig:FRitem} gives an example of this type of item).  These items are scored using a validated grading rubric that includes established point values for common responses \cite{chasteen2012cue}.  

Preliminary distractors for the CMR CUE were developed using both the FR grading rubric and student responses to the FR items.  We began by compiling a list of \emph{a priori} codes based on common responses identified on the FR rubric.  These codes were then applied to multiple semesters of student responses to the FR version.  The initial list of codes was expanded during this process to account for emergent response patterns not already encompassed by the \emph{a priori} codes.  By looking at the frequency of each code, we identified the most common justifications provided on each question.  For many of the CUE questions, a completely correct justification requires the student to connect several distinct ideas.  For example, a complete response to the item shown in Fig.\ \ref{fig:FRitem} must include several pieces: (1) that the potential can be expressed using the Multipole expansion, (2) that the cube looks like a dipole, and (3) at large distances the first non-zero term (the dipole term) will dominate.  However, both the \emph{a priori} and emergent coding showed that many students gave partially correct justifications that were missing one or more key elements.  This was especially true for the Method/Reasoning style items like the one in Fig.\ \ref{fig:FRitem}.  A standard MC format requires identifying and articulating a single, unambiguously correct answer along with three to five tempting but unambiguously incorrect distractors.  Our early attempts to construct distractors satisfying these requirements failed to capture either the variation in justifications or the partially correct ideas that were coded from the FR items.  

\begin{figure}
  \begin{minipage}{.93\linewidth}
    \vspace{-1mm}\flushleft {\bf Q3} - \emph{A solid, neutral, non-conducting cube as below, with side length `a' and $\rho(z)=kz$.  \\ \vspace{1mm}Find $\vec{E}$ or V at point P, where P is {\bf off-axis}, at a distance {\bf 50a} from the cube.\\
    \vspace{1mm} Select only one: {\bf The easiest method would be ...}}

   \begin{minipage}{0.52\linewidth}
      \vspace{1mm}\flushleft
      \emph{A. Direct Integration\\
      B. Gauss's Law\\
      C. Separation of Variables\\
      D. Multipole Expansion\\
      E. Ampere's Law\\
      F. Method of Images\\
      G. Superposition\\
      H. None of these}
   \end{minipage}
   \begin{minipage}{0.45\linewidth}    
      \begin{center}
        \includegraphics[width=19mm, height=25mm]{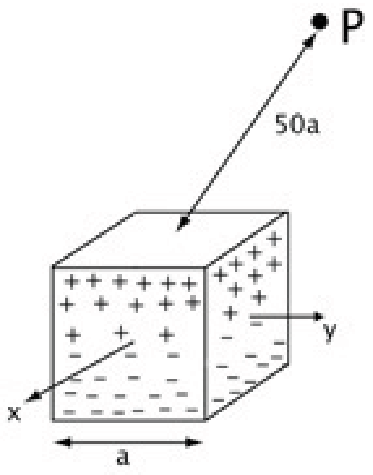}
      \end{center}
   \end{minipage}
     \flushleft{{\bf because ...} \emph{(select {\bf ALL} that support your choice of method)\\
      a. $\Box$ you can calculate $\vec{E}$ or V using the integral form \\ \hspace{4mm}of Coulomb's Law\\
      b. $\Box$ the cube will look like a dipole; approximate with \\ \hspace{4mm}$\vec{E}$ or V for an ideal dipole\\
      c. $\Box$ symmetry allows you to calculate $\vec{E}$ using a \\ \hspace{4mm}cubical Gaussian surface\\
      d. $\Box$ symmetry allows you to calculate $\vec{E}$ using a \\ \hspace{4mm}spherical Gaussian surface\\
      e. $\Box$ the observation point is {\bf far} from the cube\\
      f. $\Box$ there is not appropriate symmetry to use other \\ \hspace{4mm}methods\\
      g. $\Box$ $\nabla^2 V = 0$ outside the cube and you can solve for V \\ \hspace{4mm}using Fourier Series}} \vspace{-4pt}
  \end{minipage}
\caption{A sample item from the CMR CUE.  Question prompt has been paraphrased; see Supplemental Materials for full prompt and Ref.\ \cite{website} for rubric and scoring materials.} \label{fig:MCexample}
\end{figure}

To accommodate the wide range of correct, partially correct, and incorrect reasoning that students gave on the FR version, we switched to a less restrictive multiple-response format.  After asking the students to select the correct answer or easiest method, we provide a set of what we are calling reasoning elements, of which students can select all that support their choice of method (see Fig.\ \ref{fig:MCexample}).  Reasoning elements were taken from common codes identified in the student responses to the FR version, and each one may be correct, incorrect, or irrelevant in the context of justifying a particular answer or choice of method.  Full credit requires that a student select all (and only) the reasoning elements that together form a complete justification; however, they can receive partial credit for selecting some, but not all, of the necessary elements.   This format allows for a wider range of justifications than a standard multiple-choice test as students can combine reasoning elements in many different ways.  It also requires students to connect multiple ideas to construct a complete justification, thus providing an avenue for students to demonstrate partially correct or incomplete ideas.  

This multiple-response format is similar to the one discussed by Kubinger and Gotschall \cite{kubinger2007multipleTF}.  However, the response options they used were individually marked as true or false, making it unnecessary for the student to make any kind of connections between the statements.  In contrast, our items often include reasoning elements that are true but irrelevant, true in some physical situations, or true but incomplete.  A fully correct response with this format requires that students be consistent both between reasoning elements and method/answer selection, and between reasoning elements.  

The CMR version of the question shown in Fig.\ \ref{fig:FRitem} is given in Fig.\ \ref{fig:MCexample}.  The boxes next to each reasoning element are intended to facilitate students' interaction with the new format by resembling `check all' boxes that the students are familiar with from online surveys.  The wording of the question prompts was adjusted only when necessary to accommodate the new format.  

Roughly two thirds of the questions on the FR CUE explicitly ask for students to express their reasoning, and their CMR counterparts have the tiered format shown in Fig.\ \ref{fig:MCexample}.  The remaining items have various formats that include interpreting or generating specific formulas (e.g., boundary conditions) as well as sketching graphs, vector fields, and/or charge distributions.  These items were translated into more standard multiple-choice and multiple-response formats.  We selected the simplest format for each item that allowed us to encompass the majority of the coded student solutions to the FR version.  It is not possible to provide examples of all question formats on the CMR CUE here; however, the full instrument is included in the Supplemental Materials and can also be accessed at Ref.\ \cite{website}.

\subsection{\label{sec:Scoring}Scoring}

While allowing for a more nuanced measure of student understanding, the `select all' format of the CMR CUE also sacrifices one of the logistical advantages of standard MC questions because it is not as straightforward to score this format using automated machine grading (e.g., Scantron).  Student responses to the CMR CUE must instead be entered into an electronic grading spreadsheet.  However, once the responses have been entered, the electronic gradesheet instantly scores each student, preserving the fast and objective grading of a MC assessment.  

The new format also allows for considerable flexibility in terms of scoring.  The CMR CUE can easily be scored using multiple grading schemes simply by modifying the grading spreadsheet.  FR tests, on the other hand, require significant time and resources to regrade with a new grading scheme.  There are several different potential grading schemes for a question like the one in Fig.\ \ref{fig:MCexample} ranging from very simple to very complex.  However, Lin and Singh's previous work \cite{lin2013mcVfr} suggests that a more complex rubric designed to reflect different levels of student understanding may be more effective at achieving consistency between the FR and CMR versions.  A follow-up publication will investigate the impact of different grading schemes, but for the remainder of this paper, we will exclusively utilize a rubric (described below) designed to closely replicate the nuanced grading used to score the FR CUE \cite{chasteen2012cue}.  

The CMR CUE scoring rubric for the Method/ Reasoning questions awards full points for the easiest method, and also awards partial credit for selecting methods that are possible, even if they are not easy.  Additionally, the rubric awards points for reasoning elements that are consistent with the choice of method.   For the example item shown in Fig.\ \ref{fig:MCexample}, students are awarded three points for selecting the Multipole Expansion as the easiest method and up to a maximum of two points for any combination of the reasoning section elements `b' (1 pt), `e' (1.5 pts), and `f' (0.5 pts), for a total of five points on the question.  It is also possible, though difficult, to use Direct Integration to solve for $\vec{E}$ or V.  The rubric awards students who select method `A' one point for the Method and an additional half point for selecting the consistent reasoning element, `a'.  On items without the tiered format, students can still receive some credit for selecting distractors that reflect partially correct ideas or demonstrate an appropriate degree of internal consistency.  The point distribution and weighting of each answer or method/reasoning combination was designed to closely match the established scoring on the FR version.  

In addition to offering additional credit for consistency, the rubric also subtracts points from students with reasoning elements that are inconsistent with their choice of method.  Typically, selecting an inconsistent or incorrect reasoning element will prevent a student from getting more than three out of five points on questions that ask for explicit justifications.   On standard MC tests, a student can expect to get a score of roughly 20-25\% just by guessing.   The consistency checks in our grading scheme help to reduce the credit a student can get by chance.  Using this scoring rubric on 100 computer generated responses simulating random guessing, we found an average score of 13\%.

\subsection{\label{sec:EValidation}Expert Validation} 

The FR CUE was designed to align with learning goals for upper-division electrostatics developed in collaboration with faculty at CU.  The original instrument was also reviewed by physics experts to establish that the content was accurate, clear, and valued \cite{chasteen2012cue}.  Since the CMR CUE has the same prompts, the validity of its physics content is, to a large extent, already established.  However, the operationalization of this content has changed significantly in the new format.  We solicited and received feedback from nine experts in physics content or assessment spanning six institutions, all with experience teaching upper-division physics.  Small modifications were made to the phrasing of several items as a result of this feedback.  Overall, the expert reviewers expressed enthusiasm for the CMR CUE and offered no critiques that questioned the overall validity of the new format.  

Several of the reviewers did point out that, as with all multiple-choice formats, this new format only requires students to recognize a correct answer, which is a potentially easier task than requiring them to generate it.  Ideally, the research-generated distractors combined with the multiple-response format reduce the potential for students to simply recognize correct answers, particularly on the Method/Reasoning type questions where the individual reasoning elements rarely represent complete justifications.  In Sec.\ \ref{sec:Results} we present empirical evidence that, on average, our students do not score higher when asked to select the correct answers/justifications on the CMR version than when asked to generate answers/justifications on the FR version.

\subsection{\label{sec:Svalidation}Student Validation} 

Think-aloud validation interviews are a standard part of assessment development in order to check that students are interpreting the questions, formatting, instructions, and distractors as intended \cite{engelhardt2009ctt}.  For the CMR CUE student interviews were also crucial because we were concerned that the `select ALL that apply' format might be unfamiliar or confusing.  To date, we have performed thirteen interviews with the full 16 question CMR CUE and three interviews with a 7 question subset of the full instrument.  All interview participants had completed an upper-division electrostatics course one to four weeks prior to the interview with final course grades ranging from A to C.  During these interviews, students were asked to externalize their reasoning.  After they completed the assessment, the interviewer probed the students in more detail where it was unclear why they selected or rejected certain distractors.  

The student validation interviews were analyzed with a particular focus on identifying instances where the phrasing of an item caused students to select responses that were inconsistent with the reasoning they articulated verbally, or where students interacted with the new format in a way that caused an artificial inflation (e.g., selecting answers based on superficial similarities in wording) or deflation of their score (e.g., not following directions or not reading all the reasoning elements).  Minor wording changes were made to several of the prompts and reasoning elements as a result of these interviews.  The interviews also informed several changes to the grading scheme.  For example, some items contain reasoning elements that are true but irrelevant statements.  These were typically included because they appeared as a common justification for an incorrect method selection on the FR version.  We found in interviews that students who knew the correct method often selected these reasoning elements simply because they were true statements.  To account for this, we modified the grading rubric so students who did this would not be penalized or rewarded for selecting a true reasoning element that did not directly support their choice of method.  

A concern raised by one faculty reviewer was that students who did not know how to start a problem might figure out the correct approach by examining the given reasoning elements.  We did observe instances in the interviews where students would explicitly refer to the reasoning elements in order to inform their choice of method.  However, this technique seemed most useful to students with higher overall CUE and course scores, and, in all such cases, the student provided additional reasoning that clearly demonstrated their understanding of the correct method.  Alternatively, some students in the interviews were led down the wrong path by focusing on an inappropriate reasoning element.  This suggests that using the reasoning elements to figure out the correct method does not result in a significant inflation of scores.  

\section{\label{sec:cmrVfr}Comparing the CMR and FR CUE}

\subsection{\label{sec:Methods}Methods}

A first-order goal with the development of the CMR CUE was to achieve a meaningful level of agreement between the scores on the new CMR and well-established FR versions \cite{chasteen2012cue}.  The context for this comparison was the upper-division electrostatics course at the University of Colorado Boulder (CU).  This course, Electricity and Magnetism 1 (E\&M 1), covers the first six chapters of Griffith's text \cite{griffiths1999em} and is the first of a two semester sequence.  Typical enrollment for E\&M 1 at CU is roughly 60 physics, engineering physics, and astrophysics majors.  Data were collected during two semesters, the first of which was taught by a PER faculty member (SJP) who incorporated a number of materials designed to promote interactive engagement, such as in-class tutorials and clicker questions \cite{chasteen2012transforming}.  The second semester was taught by a visiting professor who utilized primarily traditional lecture with some minimal interactive engagement interspersed.  

To make a direct comparison of the two versions of the CUE, each semester of the E\&M 1 course was split and half the students were given the CMR version and half the FR version.  The two groups were preselected to be matched based on average midterm exam score but were otherwise randomly assigned.  Attendance on the day of the diagnostic was typical in both semesters and ultimately a total of 45 students took the CMR version and 49 students took the FR version of the CUE (75\% response rate overall).  The analysis presented in the remainder of this paper will focus exclusively on the comparisons of the FR and CMR versions of the CUE.  Future publications will report on the larger scale validation of the CMR CUE for independent implementation using data from different instructors and additional institutions.

\subsection{\label{sec:Results}Results}

\begin{figure}
    \includegraphics{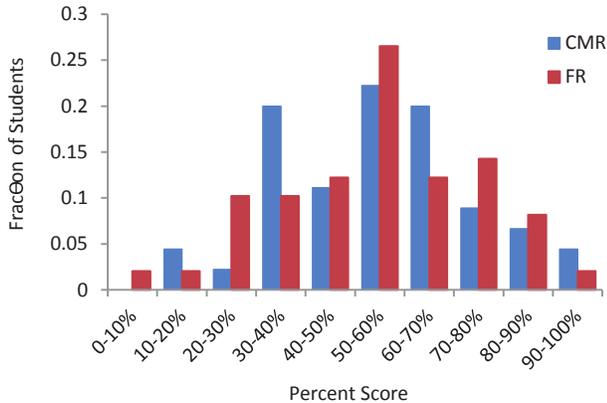}
  \caption{Distributions of scores on the CMR (N=45) and FR (N=49) CUE for the E\&M 1 course at CU.  There is no statistically significant difference between the distributions (Student's t-test, p = 0.9). }\label{fig:comparison}
\end{figure}

This section presents the quantitative comparison of test statistics from both versions of the CUE.  These test statistics are pulled exclusively from Classical Test Theory (CTT, see Sec.\ \ref{sec:Lit}).  

Using the nuanced grading rubric described in Sec.\ \ref{sec:Scoring}, the average score on the CMR version, 54.3 $\pm$ 2.8 \% ($\sigma$ = 19.1\%), was not significantly different (Student's t-test, p = 0.9) from the average on the FR version,  54.6 $\pm$ 2.8 \% ($\sigma$ = 19.6\%).  Score distributions for both versions  (Fig.\ \ref{fig:comparison}) were nearly normal (Anderson-Darling test, CMR: p = 0.9, FR: p = 0.6) and had similar variances (Brown-Forsythe test, p = 0.9).  To investigate the importance of the nuanced grading rubric to the agreement between these two versions, we also scored the CMR CUE using a strict right/wrong rubric.  In this grading scheme, students receive full credit for selecting the correct method and all the necessary reasoning elements.  Not selecting one or more of the important reasoning elements or selecting inconsistent reasoning elements both result in zero points on that item.  Using this `perfect' scoring rubric, the score on the CMR version falls to 42.8 $\pm$ 2.8 \% ($\sigma$ = 18.5\%).  The difference between this score and the average on the FR version is statistically significant (Student's t-test, p $<$ 0.01).  This finding suggests that a nuanced grading rubric designed to reflect different levels of student understanding does improve agreement between multiple-choice and free-response formats.  

The start and stop time of each student was also recorded.  On average, students spent a comparable amount of time on the CMR version, 35.0 $\pm$ 1 min ($\sigma$ = 7.5 min), as on the FR version, 34.8 $\pm$ 1 min ($\sigma$ = 7.9 min).  The average time spent on both versions was also the same for the two semesters individually.  We were initially concerned that the multiple-response format might encourage students to go through the CMR version quickly and carelessly.  Given the amount of writing required on the FR version, the fact that students took the same amount of time to complete the CMR CUE suggests that they were reading the distractors carefully and putting thought into their responses.  


\subsubsection{\label{sec:Cvalidity}Criterion Validity} 

Another property of the CMR CUE is how well its scores correlate with other, related measures of student understanding.  The most straightforward comparison is with the more traditional, long answer course exam scores.  Students in both semesters of E\&M 1 took two midterm exams and one final exam.  The CMR CUE scores correlate strongly with aggregate exam scores (Pearson Correlation Coefficient r=0.79, p $<$ 0.05).  For comparison, the correlation for the 49 students who took the FR version was also high (r = 0.79, p $<$ 0.05).  Similarly the scores for both versions are strongly correlated with final course score which includes roughly 30\% homeworks and participation (CMR: r = 0.76, FR: r = 0.73).  To account for differences between the average exam, course, and CUE scores between the two semesters, the correlations above are based on standardized scores (z-scores) calculated separately for each class using the class mean and standard deviation.  These correlations are not statistically different from the correlation, r = 0.6 (p = 0.8), reported previously for the FR CUE \cite{chasteen2012cue}.  

\subsubsection{\label{sec:Idifficulty}Item Difficulty} 

\begin{figure}
\includegraphics{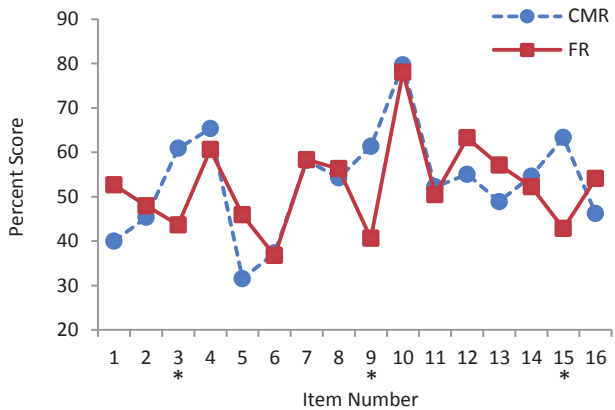}
\caption{Average scores on each item on the CUE.  Statistically significant differences between the CMR and FR versions are indicated by an asterisk (Mann-Whitney, p $>$ 0.05).  All questions are available from Ref.\ \cite{website}.} \label{fig:difficulty}
\end{figure}

In addition to looking at the overall performance of students on the CMR and FR versions of the CUE, we examined their performance on individual items.  Fig.\ \ref{fig:difficulty} shows the average scores on both versions for each question.  Differences between the scores are significant for 3 of 16 items (Mann-Whitney U-test, p $<$ 0.05; see Fig.\ \ref{fig:difficulty}).  In all three cases, the difficulty went down for the new version.  Notice that the decrease in difficulty in these three questions is balanced by a marginal (but not statistically significant) increase in difficulty on several of the remaining question; thus the whole-test average is the same for both versions.  

The FR version of two of the items with statistically significant differences (Q9 and Q15, see Supplemental Materials) were particularly challenging to adapt to the new format and, ultimately, underwent the most significant modification of all questions.  Student interviews suggest that for Q15 the decrease in difficulty arises because the FR version contains an explanation component that was eliminated on the CMR version.  Alternatively, on Q9 interviewees often recognized the appropriate justifications among the given reasoning elements even when they did not generate them which may account for the decreased difficulty.  For the remaining item (Q3, Fig.\ \ref{fig:MCexample}), we had no \emph{a priori} reason to expect that the CMR version would be significantly different than the FR version.  However, student interviews suggest that, for this item, one particularly tempting reasoning element (`b') can help students to determine the correct method (`D').  See Sec.\ \ref{sec:Reasoning} for more details on Q3.  

\subsubsection{\label{Idiscrimination}Discrimination} 

We also examined how well performance on each item compares to performance on the rest of the test (i.e., how well each item discriminates between high and low performing students).  Item-test correlations were between 0.24 and 0.71 for all items on the CMR and FR CUE with the exception of the CMR version of Q15 (r = 0.17).  Q15 was also the only item that had a statistically significant difference between the item-test correlations for the CMR and FR versions.  As stated in the previous section (Sec.\ \ref{sec:Idifficulty}), Q15 was modified significantly for the CMR version.  A common criterion for acceptable item-test correlations is r $\geq$ 0.2 \cite{ding2006bema}; however, for N = 45, correlation coefficients less than 0.24 are not statistically significant.  Q15 on the CMR CUE is the only item on either version that falls below the cutoff for acceptability or statistical significance.  Using the simpler `perfect' scoring scheme (see Sec.\ \ref{sec:Scoring}) increases the item-test correlation of Q15 to r = 0.24, above the cutoff, suggesting that a change in grading of this item may be sufficient to increase its discrimination.  

As a whole-test measure of the discriminatory power of the CMR CUE, we calculate Ferguson's Delta \cite{ding2006bema}.  Ferguson's Delta is a measure of how well scores are distributed over the full range of possible point values (total points: CMR - 93, FR - 118).  It can take on values between [0,1] and any value greater than 0.9 indicates good discriminatory power \cite{ding2006bema}. For this student population, Ferguson's Delta for both the CMR and FR versions of the CUE is 0.97.  This is similar to the previously reported FR value of 0.99 \cite{chasteen2012cue}.

\subsubsection{\label{sec:InternalC}Internal Consistency} 

The consistency of scores on individual items is also important.  To examine this, we calculate Cronbach's Alpha for both versions of the test as a whole.  Cronbach's Alpha can be interpreted as the average correlation of all possible split-half exams \cite{cortina1993ca}.    Using the point value of each item to calculate alpha, we find $\alpha$ = 0.82 for the CMR version and $\alpha$ = 0.85 for the FR version.  Again, this is consistent with the value of 0.82 reported historically for the FR CUE \cite{chasteen2012cue}.  For a unidimensional test the commonly accepted criteria for an acceptable value is $\alpha \geq$ 0.8 \cite{graham2006reliability}.  While we have no \emph{a priori} reason to assume that the CUE measures a single construct, multidimensionality will tend to drive alpha downward \cite{cortina1993ca}; thus we argue Cronbach's Alpha still provides a conservative measure of the internal consistency of the instrument.  



\subsection{\label{sec:Reasoning}Student Reasoning}

The previous section demonstrated a high degree of consistency between the CMR and FR versions on the CUE in terms of scores and a variety of test statistics.  However, one of the primary goals of the CUE's original creators was to gain insight into student thinking and the nature of common student difficulties with electrostatics \cite{chasteen2012cue}.  Gauging how much of this insight is preserved in the new CMR version requires a comparison of what students wrote/selected on each version.  To do this we performed a qualitative analysis of student responses to a subset of the CUE questions, Q1-Q7.  We focused on these seven items because they represent all the Method/Reasoning type questions (see Fig.\ \ref{fig:FRitem} \& \ref{fig:MCexample}) and typically elicit the richest and most detailed explanations on the FR version.  

We started by comparing just the students' method selections on both versions of the CUE.  This approach required coding student responses to the FR version into one of the method options offered on the CMR version.  The method coding process was relatively straightforward because the FR version directly prompts the students to select a solution method and provides them a list of methods at the beginning of the exam that matches the list provided on the CMR version.  In a few cases, some interpretation was necessary to assign a method selection to students who did not use the precise name of the method in their response (e.g., `use the multipole expansion' vs. `use a dipole approximation').  Inter-rater reliability was established by two people independently coding 20\% of the FR tests.  Agreement on the coded method selection was 96\% before discussion and 100\% after discussion.  

\begin{figure}
\includegraphics{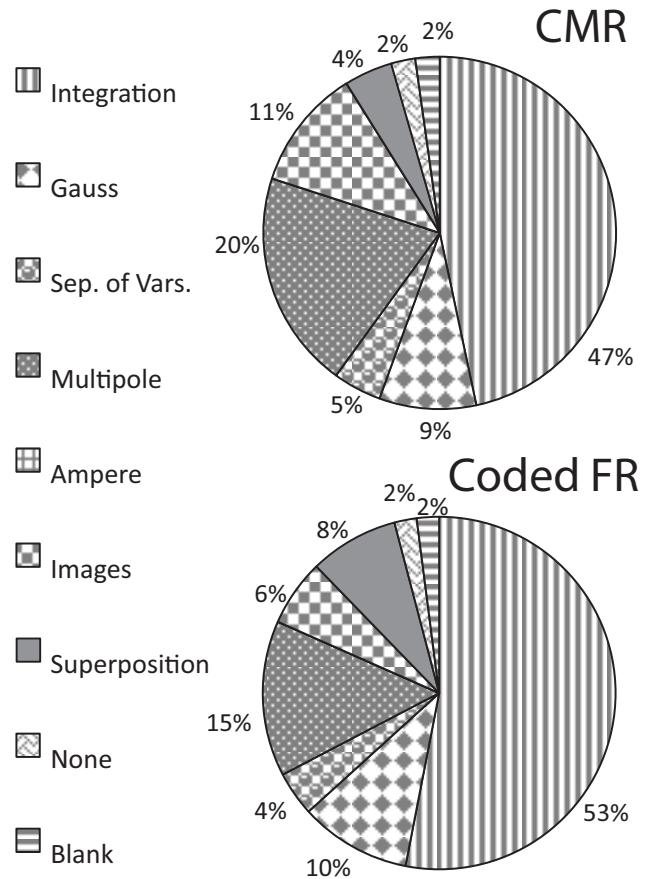}
\caption{Percent of students who selected each method on Q2 for each version of the CUE.  The top chart represents student selections from the CMR version (N=45), while the bottom chart represents coded method selections from the FR version (N=49).  The difference between the two distributions is not statistically significant (Fisher's exact test, p $>$ 0.05).} \label{fig:method1}
\end{figure}

\begin{figure}
\includegraphics{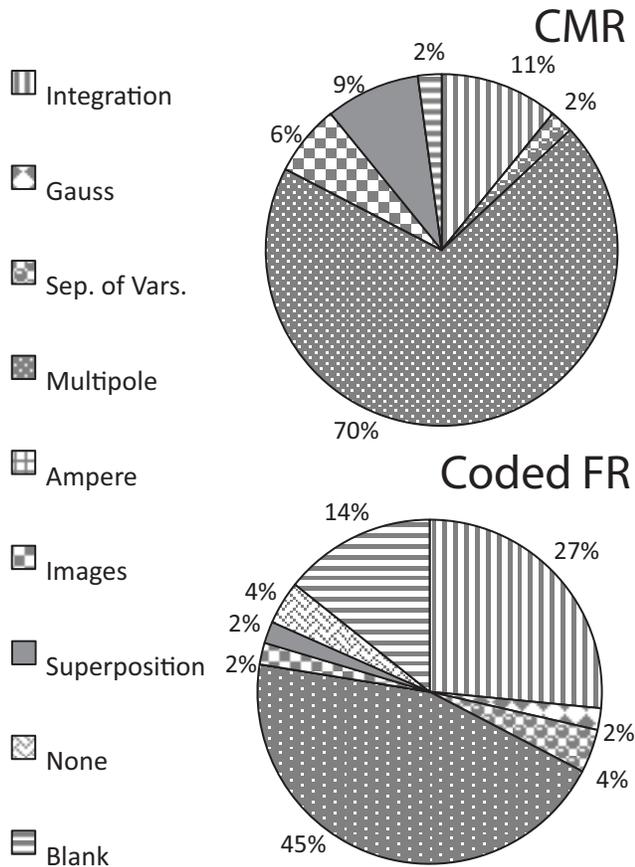}
\caption{Percent of students who selected each method on Q3 (see Fig.\ \ref{fig:MCexample}) for each version of the CUE.  The top chart represents student selections from the CMR version (N=45), while the bottom chart represents coded method selections from the FR version (N=49).  The difference between the two distributions is statistically significant (Fisher's exact test, p $<$ 0.05).} \label{fig:method2}
\end{figure}

A comparison of the method selections of students taking the CMR and FR versions of one question (Q2) is given in Fig.\ \ref{fig:method1}. Visually, the two distributions are strikingly similar, and this trend is representative of five of the seven questions.  The remaining two questions showed greater variation (see Fig.\ \ref{fig:method2}).  To quantitatively compare the two versions, we constructed 2x9 contingency tables detailing the number of students in each group that selected each method for each of the seven questions.  While $\chi^2$ is a common statistic for determining statistical significance from contingency tables, it loses considerable statistical power when cells have N $<$ 5 \cite{yates1934chiSquared}.   As many of the cells in our tables fell below this cutoff, statistical significance was determined using Fisher's Exact Test \cite{howell2012fisher, mehta1983rxcFisher}.  Fisher's Exact Test determines the probability of obtaining the observed contingency table given that the two variables in the table have no association.  It then sums the probability of the observed table along with all more extreme tables to return a p-value for having observed that particular table given the null hypothesis.

Ultimately, only the two questions with visually different distributions had statistically significant differences (p $<$ 0.05) between the method selections of students taking the CMR and FR CUE.  For one of these two questions (Q3, see Fig.\ \ref{fig:method2}), students were more likely to select the correct method (Multipole Expansion) and less likely to select the possible but harder method (Direct Integration) on the CMR version.  This trend is consistent with the decrease in difficulty observed for this item (see Sec.\ \ref{sec:Idifficulty}).  As stated earlier, this shift may be attributable to the presence of a particularly tempting correct reasoning element.  For the second of the two questions identified by Fisher's Exact Test (Q5), students were less likely to select the correct method (Superposition) and more likely to select a common incorrect method (Gauss's Law) on the CMR version.  In this case, student interviews suggest that this trend may be due to the presence of a particularly tempting but this time incorrect reasoning element justifying the use of Gauss's Law.  

The reasoning portion of the FR questions is more challenging to code than the method portion because students are no longer constrained to a finite list of methods and are free to justify their answer in any way they choose.  We started by coding students' free responses using the reasoning elements provided on the CMR version.  We also identified aspects of student responses that did not match one of the available reasoning elements.  These aspects were coded into two `other' categories: satisfactory and unsatisfactory.  Satisfactory codes were given to elements of students' justifications that represented correct physical statements that supported the choice of method but that did not get coded into one of the CMR categories.  Unsatisfactory codes were given to elements that represented incorrect or irrelevant statements.  Students could receive multiple codes, meaning that a student could be awarded an `other' code even if some elements of their response fit into one of the CMR categories.  

Due to the higher degree of difficulty inherent in coding the reasoning portion, inter-rater reliability was established in two stages.  Additionally, because the coding on the reasoning portion allows for multiple codes for each student, we determined inter-rater reliability statistics for both complete agreement (i.e., no missing or addition codes) and partial agreement (i.e., at least 1 overlapping code).  First stage reliability statistics were generated from independent coding of 20\% of the FR exams.  For this initial set, complete agreement was 65\% before discussion and 94\% after discussion, and partial agreement was 79\% before discussion and 96\% after discussion.  In the second stage, an additional 10\% of the FR exams were independently coded, and both complete and partial agreement before discussion rose to 89\%.  There is not a well accepted threshold for an acceptable percent agreement because this cutoff must account for the possibility of chance agreement and thus depends on the number of coding categories \cite{lombard2002reliability}.  However, given our large number (7-10) of non-exclusive coding categories, the potential for complete agreement by chance is low.  Thus, we consider 89\% agreement to be acceptable for the general comparisons made here.  

Ultimately, an average of 74\% of FR students who did not leave the reasoning portion blank were coded as having one or more of the CMR reasoning elements per question.  The remaining 26\% received only `other' codes (11\% satisfactory and 15\% unsatisfactory) meaning that no aspect of their justification matched one of the CMR reasoning elements.  Overall, 33\% of students who took the FR version received `other' codes, including those who also received one or more CMR codes.  In other words, one third of the information on students' reasoning that is accessed by the FR version is forfeit on the CMR version.  This result is not surprising as it is not possible to capture the full range of student reasoning with a finite number of predetermined response options.  

This section presented an analysis of student responses to the Method/Reasoning type questions on both versions of the CUE.  We found that the two versions elicited matching Method selections for five of seven questions.  On the remaining two questions, there was a statistically significant shift in the fraction of students selecting each of the two most common method choices.  In both cases, this shift may be attributable to the presence of a particularly attractive reasoning element.  Additionally, we find that roughly three-quarters of responses to the FR version contained elements that matched one or more of the reasoning options provided on the CMR CUE.  However, roughly a third of these responses also contained elements that did not match the CMR reasoning options; thus, the logistical advantages of our CMR assessment come at the cost of reduced insight into student reasoning.

\section{\label{sec:Discussion}Summary and Discussion}

We have created a novel multiple-response version of an existing upper-division conceptual assessment, the CUE.  Using student responses to the original free-response version of the instrument, we crafted multiple-response options that reflected elements of common student ideas.  This new version utilizes a novel approach to multiple-choice questions that allows students to select multiple reasoning elements in order to construct a complete justification for their answers.  By awarding points based on the accuracy and consistency of students' selections, this assessment has the potential to produce scores that represent a more fine-grained measure of students' understanding of electrostatics than a standard multiple-choice test.  

Direct comparison of the multiple-response and free-response versions of the CUE from two large, upper-division electrostatics courses yielded the same average score when using a nuanced grading scheme on both.   The two versions also showed a high degree of consistency on multiple measures of test validity and reliability.  Student interviews and expert feedback were used to establish the content validity of the CMR CUE for this comparison.  Given the agreement between scores on the two versions and the ease of grading afforded by this new format, the CMR CUE is a considerably more viable tool for large-scale implementation.  Additionally, while the FR version elicits greater variation in student reasoning, nearly three-quarters of students' responses to the FR version contain one or more elements that match the reasoning elements provided on the CMR version.  

Our findings suggest that the CMR format can provide easily-graded questions that produce scores that are consistent with scores from a FR format.  We found this outcome surprising.  When we began developing the CMR CUE we were skeptical that it would be possible to maintain more than a superficial level of consistency between the two versions.  However, construction of the reasoning elements for the CMR CUE items relied heavily on the existence of data from student responses to the FR version.  It is our opinion that, without this resource, our CMR CUE would not have been as successful at matching the FR CUE.  An important limitation of the CMR format may be its reliance on pre-existing data from a FR version of the item or test.  Additionally, as with the majority of conceptual assessments, the CMR CUE took several years to develop and validate even when building off the already established FR CUE.  This time requirement places significant constraint on the development of similar assessments by instructors.  

Another potential limitation of the CMR format comes from the relative complexity of the prompts.  It is important that students read the directions fully for each question in order for an instructor to meaningfully interpret their response patterns.  Interviews and overall scores from the two electrostatics courses discussed here suggest that our students followed directions and engaged with the question format as expected.  However, these students were all junior and senior level physics students taking a course required for their major.  More experimentation is necessary to determine if the CMR format is viable for use with less experienced students who are not as invested in the content (e.g., introductory students or non-majors).  

Additionally, not all questions easily lend themselves to the CMR format.  For example, Q9 on the CUE (see Supplemental Materials) was particularly challenging to translate into a CMR format.  Q9 deals with determining the sign of the electric potential from a localized charge distribution given an arbitrary zero point.  Students can leverage multiple, equally valid ways of determining the correct answer (e.g., by thinking about work, electric field, or shifting the potential).  Capturing all of the correct, incorrect, and partially correct ideas expressed by students on the FR version of this question would have required a prohibitively large number of reasoning elements.  To avoid this, we crafted a smaller number of reasoning elements with the goal of including each of the different types of justification (i.e., work, electric field, potential); however, we recognized that these elements did not encompass the variety of partially correct or incomplete ideas present in the FR version.  

This paper has focused exclusively on a direct comparison of the CMR and FR versions of the CUE in order to establish the extent to which the two formats are consistent.  Ongoing work with the CMR CUE will include building on the analysis presented here to more robustly establish its validity and reliability as an independent instrument.  To do this, we are expanding data collection with an emphasis on including additional instructors at multiple institutions.  Targeting multiple instructors will help us to determine if the CMR CUE retains the FR version's sensitivity to differences in pedagogy.  See Ref.\ \cite{website} for more information on reviewing or administering the CMR CUE. 

\begin{acknowledgments}
Particular thanks to the PER@C group and the faculty who reviewed the CUE.
This work was funded by NSF-CCLI Grant DUE-1023028 and an NSF Graduate Research Fellowship under Grant No. DGE 1144083.
\end{acknowledgments}

\bibliography{master-refs}
\bibliographystyle{apsper}   

\end{document}